\documentclass{article}
\usepackage[a4paper, total={7in,10in}]{geometry}
\usepackage{hyperref}
\usepackage{epsfig,amssymb,amsmath,textcomp,caption,verbatim,gensymb,braket}
\usepackage{epstopdf,subfig}
\usepackage[normalem]{ulem}
\usepackage{graphicx}
\usepackage{amsfonts}
\usepackage{titling,multicol,float}
\begin{document}
		
	\title{An attractive spin-orbit potential from the Skyrme model: erratum}
	
	\author{Chris Halcrow and Derek Harland\\
	School of Mathematics, University of Leeds}
	
	\date{7th August 2022}
	
	\maketitle
	
\begin{multicols}{2}
	
Since publication we have noticed two errors in our article (which follows this erratum).  The first is in our calculation for the tensors $A_{ij;ab}$ etc, which was based on a calculation in \cite{schroers1993}.  In both our paper and \cite{schroers1993}, an acceleration was incorrectly neglected.  The correct expressions for these tensors are
\begin{align*}
A_{ij;ab}&=\varepsilon_{ajc}\left[
-\delta_{ib}\nabla_c \left(\frac{e^{-m_\pi X/\hbar}}{X}\right)
-\nabla_{ibc}\left(\frac{\hbar e^{-m_\pi X/\hbar}}{2m_\pi}\right)
\right]  \\ 
B_{ij;ab}&= -\varepsilon_{aic}\varepsilon_{bjd}\nabla_{cd}\left(\frac{\hbar e^{-m_\pi X/\hbar}}{m_\pi}\right) \\
C_{ij;ab}&=\delta_{ij}\nabla_{ab}\left(\frac{e^{-m_\pi X/\hbar}}{2X}\right)
-\nabla_{abij}\left(\frac{\hbar e^{-m_\pi X/\hbar}}{4m_\pi}\right) \\&
-(\delta_{jb}\nabla_{ia}
+\delta_{ja}\nabla_{ib}
+\delta_{ib}\nabla_{ja}
+\delta_{ia}\nabla_{jb})\left(\frac{3e^{-m_\pi X/\hbar}}{8X}\right)
\\
D_{ab}&=\nabla_{ab}\left(\frac{e^{-m_\pi X/\hbar}}{X}\right).
\end{align*}
There is also an error in the formula for the hamiltonian $\mathcal{H}$ that precedes eq.\ (5), but this does not affect the calculation of the spin-orbit potential.  For more details on both errors, see \cite{HH2021}.

As a result of the first error, eq.\ (14) is incorrect.  The correct formula for the spin-orbit potential is
\begin{align*} \label{H22}
 H_{LS}^2 = \,&\frac{\rho^2e^{-2s} }{972 \hbar^3 M X^8\Lambda^2} \boldsymbol{L}\cdot (\boldsymbol{\sigma}_1 + \boldsymbol{\sigma}_2)\Big( 64 \Lambda^4 (s^2+3s+3)^2 \nonumber \\
 &-32 \hbar^2X^2(s+1)(7s^2-6s-6) \nonumber
 \\ &+\hbar^4 X^4(s+1)(115 s-101) \Big).
 \end{align*}
This hamiltonian still allows for a negative spin-orbit potential: for example, when $m_\pi=0$ the leading large $r$ behaviour is $H_{LS}^2 \sim -101\hbar\rho^2/972M X^4\Lambda^2$.  However, with the calibration used in the paper (originally proposed by Lau--Manton) the spin-orbit potential is positive at intermediate separations $r\sim 2\text{fm}$, as shown in Fig.\ \ref{fig:new}.  In a forthcoming paper, we show that the correct sign is obtained at intermediate separations using a more sophisticated approximation based on instanton holonomies \cite{HHinstantons}.

\begin{figure}[H]
\begin{center}
\includegraphics[width=0.45\textwidth]{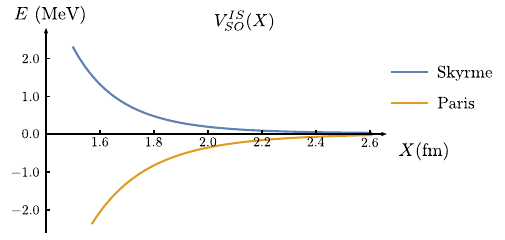}
\end{center}
\caption{A comparison between the isoscalar spin-orbit force from the Skyrme and phenomenological Paris potential.  This replaces figure \ref{fig:old} in the paper below.}
\label{fig:new}
\end{figure}
\addtocounter{figure}{-1}

\end{multicols}
	
	\newpage

	\title{An attractive spin-orbit potential from the Skyrme model}
	
	\author{Chris Halcrow and Derek Harland\\
	School of Mathematics, University of Leeds, Leeds, LS2 9JT, United Kingdom}
	
	\date{2nd July 2020}
	
	\maketitle
	
\begin{abstract}
We derive the nucleon-nucleon isoscalar spin-orbit potential from the Skyrme model and find good agreement with the Paris potential.  This solves a problem that has been open for more than thirty years and gives a new geometric understanding of the spin-orbit force.  Our calculation is based on the dipole approximation to skyrmion dynamics and higher order perturbation theory.
\end{abstract}
	
%	\keywords{spin-orbit, skyrmion, nucleon, qcd}

\begin{multicols}{2}
Understanding the nucleon-nucleon interaction is a fundamental and challenging problem. Even 85 years after Yukawa's pioneering work \cite{Yukawa1935}, our knowledge of the short-range proton-neutron interaction is essentially phenomenological. The spin-orbit force, which favours nucleon-nucleon configurations where the relative orbital angular momentum of the nucleons is aligned with the sum of their spins, was first studied by Signell and Marshak and by Gammel and Thaler \cite{SM1958,GT1957}. Without the force, nucleon-nucleon scattering data cannot be reproduced and the correct nuclear magic numbers cannot be found \cite{Mayer1948}.

To understand the nuclear force from first principles one should study QCD. Unfortunately the theory is non-perturbative at low energies, making a first principles calculation prohibitively difficult. Instead, an effective theory such as Chiral Effective Field Theory must be used \cite{Weinberg1978}.  Here, the quarks and gluons are ``integrated out" leaving the hadrons such as pions, kaons and nucleons, acting as the fundamental particles. Unfortunately, every new term included in the Lagrangian comes with at least one new parameter. The problem gets worse as more fields are added; not only do their kinetic contributions arrive with parameters, so do their couplings with every other particle in the theory. The proliferation of parameters limits the predictive power of the theory.

The $SU(2)$ Skyrme model is closely related to Chiral Effective Field Theories but the only fundamental field is the pion. Skyrme realised that the basic pion theory had an interesting mathematical structure: one which allowed for the creation of topologically non-trivial pion field configurations, now called skyrmions \cite{Skyrme1961}. Such fields have an integer-valued conserved charge called the topological charge. Skyrme identified this integer with the nucleon number and skyrmions with nucleons. In this way nucleons are not added as new fields, but are constructed from the pion fields and no additional parameters are needed to describe nucleons.  The model is now understood to be a large-$N_C$ description of QCD \cite{witten1983b}, and has links to holographic QCD \cite{sakaisugimoto2005}.

The one-nucleon sector was first studied in \cite{ANW1983}, and the quantised skyrmion gives a good description of the nucleon. The study of the nucleon-nucleon potential in the Skyrme model has a long history.  It was realised early on that the Skyrme model successfully reproduces the one-pion exchange potential \cite{JJP1985,VMLLCL1985}.  The central potential is also successfully reproduced, but only in calculations that include higher order corrections in perturbation theory \cite{WAH1992}.  However, calculations of the isoscalar spin-orbit potential consistently produced a potential with the wrong sign \cite{OSYKS1988,RD1988,ASW1993}.  A proposal in \cite{RS1989} that adding a sextic term to the Lagrangian would change this result was eventually refuted \cite{Abada1996}.  Better results were obtained in \cite{KE1995,Abada1997}, but only at the expense of including additional fields in the model. The lack of a simple, positive result for the spin-orbit potential has been the major shortcoming in the Skyrme model's description of the nucleon-nucleon interaction. But these papers have one thing in common: they are all based on the product approximation.  This approximation is now recognised to be unreliable except at large separations (see e.g. \cite{VWWW1987}).  We will argue later that the product approximation is to blame for these historical negative results.

In this letter we calculate the spin-orbit potential using a new method that is inspired by a geometrical understanding of the spin-orbit force.  Our calculation is based on higher order perturbation theory and the dipole approximation.  The dipole approximation is valid at large separations in many variants of the Skyrme model, including those recently developed to improve the binding energy of skyrmions \cite{ANSGW2013,GHS2015,sutcliffenaya2018}, so our results are robust and widely applicable. Our method could also be adapted for use in holographic QCD. For the standard Skyrme model, we show that the potential matches the phenomenologically successful Paris potential. Overall, we show that the Skyrme model does reproduce the isoscalar spin-orbit force essential for nuclear physics and gives a new, geometric interpretation of its origins.

The  fundamental field of the Skyrme model is $U \in SU(2)$, written in terms of the pion fields $\boldsymbol{\pi}$ as
\begin{equation*}
U(\boldsymbol{x}) = \begin{pmatrix} \pi_0+i\pi_3 & i\pi_1 +  \pi_2 \\ i\pi_1 - \pi_2 & \pi_0 - i\pi_3 \end{pmatrix}  ,
\end{equation*}
where $\pi_0$ is an auxilary field satisfying $\pi_0^2 + \boldsymbol{\pi}\cdot \boldsymbol{\pi} = 0$. The standard Skyrme Lagrangian is
\begin{align} \label{Lagrangian}
\mathcal{L}
= 
&\text{Tr}  \Big( \frac{F_\pi^2}{16\hbar} \partial_\mu U \partial^\mu U^\dagger +\frac{\hbar}{32e^2} (\partial_\mu U \partial_\nu U^\dagger - \partial_\nu U \partial_\mu U^\dagger) \nonumber \\ &\times(\partial^\mu U \partial^\nu U^\dagger - \partial^\nu U \partial^\mu U^\dagger) 
-\frac{F_\pi^2m_\pi^2}{8\hbar^3}(1-U) \Big) , 
\end{align} 
where $m_\pi$ is the pion mass, $F_\pi$ is the pion decay constant and $e$ is a dimensionless parameter. A $B$-skyrmion is a static solution to the equations of motion of \eqref{Lagrangian} with topological charge $B$.

A 1-skyrmion, which models a nucleon, is parameterised by a position and an orientation. The skyrmion-skyrmion system (in the zero center of mass frame) can then be described using a configuration space parameterised by a relative position $\boldsymbol{X}$ and two $SO(3)$-valued orientation matrices $R_1$ and $R_2$. We also define the relative orientation matrix as $R = R_1^{-1}R_2$. The long-range potential energy between two skyrmions is well known and reproduces the one-pion exchange potential
\begin{equation*}
V(R,\boldsymbol{X}) = 2\rho R_{ab}\nabla_{ab} \left(e^{-m_\pi X/\hbar}/X\right), 
\end{equation*} 
where $\rho=8\pi\hbar^3 C_1^2/e^4F_\pi^2$ and $\nabla_{ab} = \nabla_a\nabla_b$ acts on $X = |\boldsymbol{X}|$. The dimensionless constant $C_1$ is determined by the asymptotic behaviour of the pion fields of a $1$-skyrmion \cite{Feist2011}. The interaction potential depends on the relative orientation: it is most attractive when one skyrmion is rotated by $\pi$ around an axis perpendicular to the line joining it to the other. This is called the attractive channel. Using numerical techniques we can solve the equations of motion to see how separated skyrmions evolve in the attractive channel \cite{Verbaarschot1987} (see Figure 1). The skyrmions begin to merge as they approach and eventually form a torus.  This torus is the 2-skyrmion and represents the point of closest approach; if we were to continue the simulation the skyrmions would re-emerge at right angles to their path of approach. The skyrmions cannot get too close; not because of a repulsive short range potential but instead due to the geometry of the $2$-skyrmion configuration space.

\begin{figure}[H]
\begin{center}
	\includegraphics[height=0.18\textwidth]{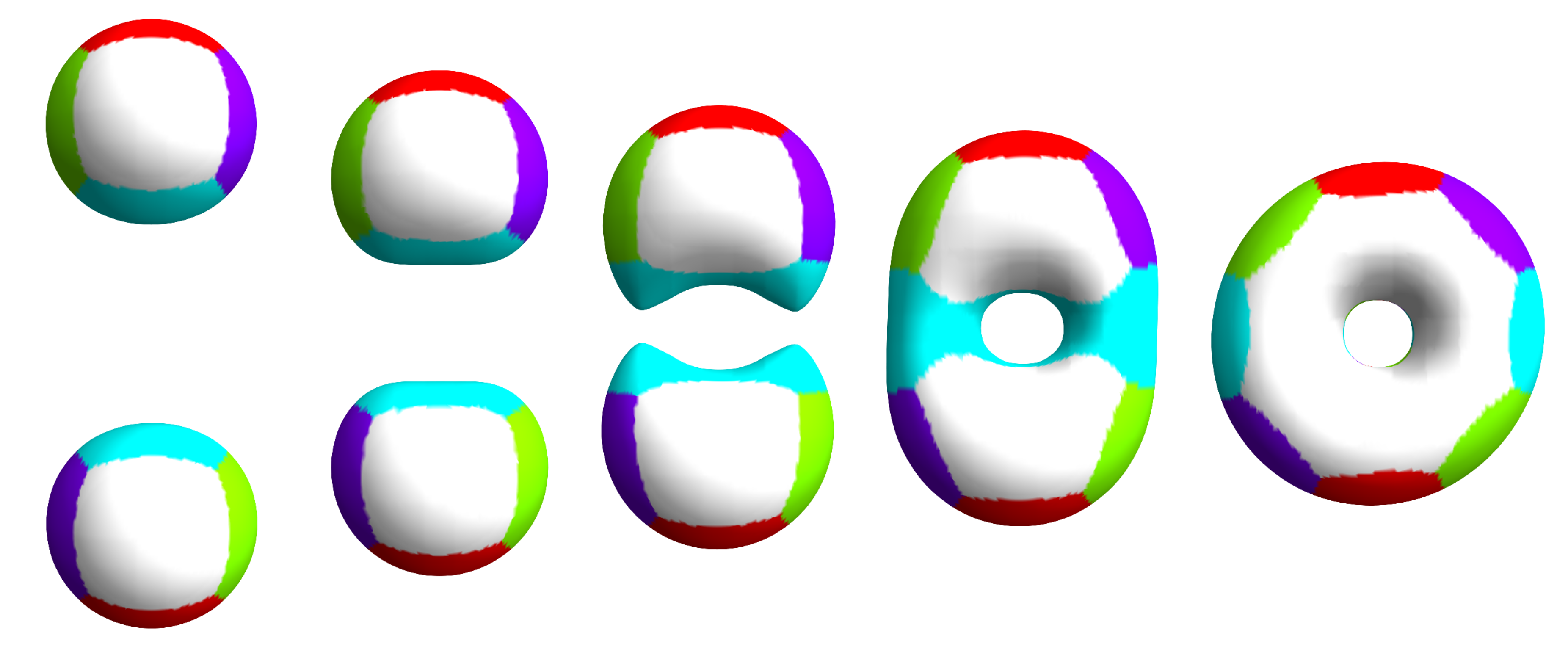}
	\end{center}
	\caption{Two skyrmions interact in the attractive channel. Starting from a large separation, the skyrmions attract and form a torus. The colouring indicates orientation, as in \cite{GHS2015}.}
\end{figure}

The torus has more symmetry than a generic $2$-skyrmion configuration and this has consequences for the spin-orbit force. Consider the configurations in Figure 1 and the following transformation: rotate the entire system around the $z$-axis (facing the reader) by $\theta$, then rotate each skyrmion around its own $z$-axis by $-2\theta$. This is a continuous path on the configuration space. Along this path, the orbital angular momentum of the skyrmions and their spins are anti-aligned. If the transformation is applied to the torus, nothing happens. It is a symmetry of the configuration. Hence this path has zero length at the torus, and is short nearby. So, paths where the spin and orbital angular momentum are anti-aligned are shorter than one naively expects. In quantum mechanics, short paths imply high energy. For example, the energy of a free particle in a $1$D box scales with the inverse square of the box length. Hence, wavefunctions with spin and orbital angular momentum anti-aligned have high energy. This is exactly the consequence of the spin-orbit force. The argument gives a geometric understanding of the force: it is ultimately due to the preference for the attractive channel and existence of a toroidal $2$-skyrmion. 

To see if our geometric intuition does generate the expected spin-orbit force we must calculate the effective nucleon-nucleon hamiltonian from the Skyrme model. We start by considering the asymptotic interaction of skyrmions in the center of mass frame. When widely separated, we can write the Lagrangian as $\mathcal{L} = \mathcal{L}_0 + \rho \mathcal{L}_\text{int}$, with
\begin{equation} \label{Lfree}
\mathcal{L}_0 = \tfrac{M}{4} \dot{\boldsymbol{X}}^2+ \tfrac{\Lambda}{2} \boldsymbol{\omega_1}^2 + \tfrac{\Lambda}{2} \boldsymbol{\omega_2}^2 \, .
\end{equation}
Here $M$ and $\Lambda$ are the mass and moment of inertia of the 1-skyrmion, and $\boldsymbol{\omega}_\alpha$ are angular velocities, defined by $R_\alpha^{-1}\dot{R_\alpha}=\boldsymbol{\omega}_\alpha\cdot\boldsymbol{J}$ with $J_i$ satisfying $[J_i,J_j]=\epsilon_{ijk}J_k$. The interaction Lagrangian has been studied in the case $m_\pi=0$ \cite{schroers1993}, and we follow this approach to find the result with $m_\pi \neq 0$. Far from its center, each skyrmion looks like a triplet of dipoles with dipole moments $\boldsymbol{p}_a = 4\pi C_1\boldsymbol{e}_a$, where $\boldsymbol{e}_a$ are a triplet of orthonormal vectors \cite{schroers1993}. We can then apply the theory of relativistic dipoles to find the interaction Lagrangian. We will consider low energy nucleon-nucleon interactions and hence, like in \cite{schroers1993}, we neglect terms with more than two time derivatives. In this way, we can find an interaction Lagrangian which depends on the relative separation, orientation matrices and angular velocities $\boldsymbol{\omega}_i$ of the skyrmions. It is
\begin{equation} \label{Lint}
\mathcal{L}_\text{int} = \rho\Big(\dot{X}^iA_{ij}\omega_1^j+\dot{X}^iA_{ij}^*\omega_2^j +
 \omega_1^iB_{ij}\omega_2^j + \dot{X}^iC_{ij}\dot{X}^j - 2D \Big)
\end{equation}
with, for example, $A_{ij} = A_{ij;ab}R_{ab}$ and 
\begin{align*}
A_{ij;ab} &=  \epsilon_{ajc}\left[ \delta_{ic}\nabla_b\left(\frac{ e^{-m_\pi X/\hbar}}{X}\right)  - \nabla_{icb}\left( \frac{\hbar e^{-m_\pi X/\hbar}}{2m_\pi} \right)\right]\\
B_{ij;ab} &= -\epsilon_{aic}\epsilon_{bje}\nabla_{ce}\left( \frac{\hbar e^{-m_\pi X/\hbar}}{m_\pi} \right) \\
C_{ij;ab} &=  \delta_{ij}\nabla_{ab}\left(\frac{e^{-m_\pi X/\hbar}}{2X}\right) - \nabla_{abij}\left(\frac{\hbar e^{-m_\pi X/\hbar}}{4m_\pi}\right)+ \\ &
 (\delta_{jb}\nabla_{ia} + \delta_{ja}\nabla_{ib} + \delta_{ib}\nabla_{ja} + \delta_{ia}\nabla_{jb})\left(\frac{e^{-m_\pi X/\hbar}}{8X}\right) \\
D_{ab}  &= \nabla_{ab}\left(\frac{e^{-m_\pi X/\hbar}}{ X}\right) \, ,
\end{align*}
and $A^*_{ij} = A_{ij;ba}R_{ab}$. The result is similar to the result \cite{schroers1993} for massless pions, and agrees in the limit $m\to 0$. We can write the total Lagrangian in terms of a metric
\begin{align} \label{Lmet}
\mathcal{L} = &\tfrac{1}{2}(\dot{\boldsymbol{X}}, \boldsymbol{\omega}_1, \boldsymbol{\omega_2})^T\cdot( g + \delta g ) \cdot(\dot{\boldsymbol{X}}, \boldsymbol{\omega}_1, \boldsymbol{\omega_2}) - 2\rho D
\end{align}
where $g$ and $\delta g$ are functions of $\boldsymbol{X}$ and $R$ which can be read off from equations \eqref{Lfree} and \eqref{Lint}. We will assume that $X$ is large and treat $\delta g$ as a small perturbation of $g$.

The Lagrangian (4) can be quantised by a standard method. The quantum hamiltonian $\mathcal{H}$ is written in terms of the Laplace-Beltrami operator $\Delta_{g+\delta g}$, in turn written in terms of the vector fields $\boldsymbol{E} = ( i\boldsymbol{P}/\hbar, -i\boldsymbol{S}_1,-i\boldsymbol{S}_2)$ which are dual to $( \dot{\boldsymbol{X}}, \boldsymbol{\omega}_1, \boldsymbol{\omega}_2)$. Evaluating the known expression for $\mathcal{H}$ in powers of $\delta g$, we find that
\begin{align*} \label{Hexp}
&\mathcal{H} = |1+g^{-1}\delta g|^{1/4}\left( -\frac{\hbar^2}{2}\Delta + V\right) |1+g^{-1}\delta g|^{-1/4}  \\ &= \frac{\hbar^2}{2}\big(-\Delta_g + E_\kappa ( g^{\kappa\lambda} \delta g_{\lambda\mu} g^{\mu \sigma} - g^{\kappa\lambda} \delta g_{\lambda\mu} g^{\mu \nu} \delta g_{\nu \rho} g^{\rho\sigma}) E_\sigma \nonumber \\
 &+\frac{1}{16}g^{\mu\nu}[E_\mu,g^{\kappa\lambda}\delta g_{\lambda\kappa} ] [E_\nu,g^{\rho\sigma}\delta g_{\sigma\rho}]+O(\delta g^3) \big)+ 2\rho D \nonumber \, 
\end{align*}
(here conjugation with the determinant $|1+g^{-1}\delta g|^{1/4}$ ensures that $\mathcal{H}$ is hermitian w.r.t.\ $g$).  Evaluating this expression using $g$ and $\delta g$ derived from \eqref{Lfree} and \eqref{Lint}, we find
\begin{equation} \label{Hfree}
\mathcal{H} = \frac{1}{M} |\boldsymbol{P}|^2 + \frac{\hbar^2}{2\Lambda}|\boldsymbol{S}_1|^2 + \frac{\hbar^2}{2\Lambda} |\boldsymbol{S}_2|^2 + \rho\mathcal{H}_I
\end{equation}
with
\begin{align}\label{HI}
\rho\mathcal{H}_I =  &\frac{\rho\hbar}{M\Lambda}(P^iA_i+A_i^\dagger P^i) - \frac{\rho^2\hbar}{M\Lambda^2}(P^i\tilde{A}_i+\tilde{A}_i^\dagger P^i) \nonumber \\ &+ 2\rho \tilde{D} + \rho^2F + O(\rho^3)+O(M^{-2})
\end{align}
and
\begin{align*}
A_i &= A_{ij}S_1^j+A^\ast_{ij}S_2^j, \quad \tilde{A}_i = A^\ast_{ij}B_{kj}S_1^k + A_{ij}B_{jk}S_2^k \\
\tilde{D} &= D -\frac{\hbar^2}{4\Lambda^2}(S_1^iB_{ij}S_2^j+S_2^jB_{ij}S_1^i) \\
F &= \frac{\hbar^2}{2\Lambda^3}(S_1^iB_{ij}B_{kj}S_1^k+S_2^iB_{ji}B_{jk}S_2^k)+\frac{\hbar^2}{\Lambda^2M}A_i^\dagger A_i.
\end{align*}

When the separation $X$ is large and momentum $P$ is small (true for low energy scattering), the most important part of the hamiltonian \eqref{Hfree} is $\mathcal{H}_0=\frac{\hbar^2}{2\Lambda}(|\boldsymbol{S}_1|^2+|\boldsymbol{S}_2|^2)$.  The eigenvalues of this operator are of the form $\frac{\hbar^2}{2\Lambda}(\ell_1(\ell_1+1)+\ell_2(\ell_2+1))$ for positive half-integers $\ell_1,\ell_2$.  The eigenspace for the lowest eigenvalue $3\hbar^2/4\Lambda$ describes two particles with spin $\frac12$ and isospin $\frac12$ i.e.\ a pair of nucleons.  The low-energy dynamics of the hamiltonian \eqref{Hfree} can be described using an effective hamiltonian acting on this eigenspace.  This effective hamiltonian can be calculated using degenerate perturbation theory: for a hamiltonian of the form $\mathcal{H}_0+\delta\mathcal{H}$ the formula is
\begin{multline*}
E_0 + \delta\mathcal{H}^{00} - \sum_{N>0} \frac{\delta\mathcal{H}^{0N}\delta\mathcal{H}^{N0}}{E_N-E_0} \\
+\sum_{M,N>0}\frac{\delta\mathcal{H}^{0N}\delta\mathcal{H}^{NM}\delta\mathcal{H}^{M0}}{(E_N-E_0)(E_M-E_0)}  \\
 -\sum_{N>0} \frac{\delta\mathcal{H}^{0N}\delta\mathcal{H}^{N0}\delta\mathcal{H}^{00}+\delta\mathcal{H}^{00}\delta\mathcal{H}^{0N}\delta\mathcal{H}^{N0}}{2(E_N-E_0)^2} + O(\delta\mathcal{H}^4). 
\end{multline*}
Here $E_0<E_1<\ldots$ are the eigenvalues of $\mathcal{H}_0$, and $\delta\mathcal{H}^{NM}$ maps from the $E_M$-eigenspace to the $E_N$-eigenspace so that $\delta\mathcal{H}=\sum_{M,N} \delta\mathcal{H}^{NM}$.  In the situation at hand, with $\delta\mathcal{H}=M^{-1}|P|^2+\rho \mathcal{H}_I$, the formula gives
\begin{align} \label{EH}
\mathcal{H}_\text{eff} = E_0 + \frac{|P|^2}{M}+\rho\mathcal{\mathcal{H}}_I^{00} - \rho^2\sum_{N>0}\frac{\mathcal{H}_I^{0N}\mathcal{H}_I^{N0}}{E_N-E_0} \nonumber \\
+ \frac{\rho^2}{2M}\sum_{N>0}\frac{\mathcal{H}_I^{0N}[|P|^2,\mathcal{H}_I^{N0}] - [|P|^2,\mathcal{H}_I^{0N}]\mathcal{H}_I^{N0}}{(E_N-E_0)^2} \nonumber \\ + O(\rho^3) + O(M^{-2}).
\end{align}
The hamiltonian \eqref{EH} is necessarily of the same form as the nucleon-nucleon potential, because the skyrmion-skyrmion and nucleon-nucleon systems enjoy the same symmetries.  We will only calculate the isoscalar spin-orbit term, which takes the form
\begin{equation}
H_{LS}=\frac{1}{\hbar}V_{LS}^{IS}\big(X,|P|^2,|L|^2\big)\,\boldsymbol{L}\cdot(\boldsymbol{\sigma}_1+\boldsymbol{\sigma}_2),
\end{equation}
where $\boldsymbol{\sigma}_i$ are the spin Pauli matrices. $H_{LS}$ can be expressed as a power series in $\rho$ and $M^{-1}$, and we only calculate the leading term, which occurs at order $\rho^2M^{-1}$.

First we evaluate the contribution to $H_{LS}$ from the first order term $\rho\mathcal{H}_I^{00}$ from \eqref{EH}.  Here the only term in \eqref{HI} which can contribute to $H_{LS}$ is the term which is linear in momentum.  We thus obtain a contribution
\begin{align}\label{H21}
&H_{LS}^1 = \frac{\rho\hbar}{M\Lambda} P_i(A_i+A_i^\dagger)^{00} - \frac{\rho^2\hbar}{M\Lambda^2}P^i(\tilde{A}_i+\tilde{A}_i^\dagger)^{00} \\
&= -\frac{\rho^2\hbar e^{-2s}}{3M\Lambda^2 X^4}\boldsymbol{L}\cdot(\boldsymbol{\sigma}_1+\boldsymbol{\sigma}_2)\left(3+4s + s^2\right) \nonumber
\end{align}
where $s=m_\pi X/\hbar$.  This result is obtained using two projection theorems, both of which can be derived from eq.\ \eqref{RMN} below (see also \cite{Abada1996,harlandmanton}): $R_{ab}^{00}=\sigma_{1a}\sigma_{2b}\boldsymbol{\tau}_1\cdot\boldsymbol{\tau}_2/9$ and $(R_{ab}R_{cd})^{00}=\delta_{ac}\delta_{bd}/3+\varepsilon_{ace}\varepsilon_{bdf}\sigma_{1e}\sigma_{2f}\boldsymbol{\tau}_1\cdot\boldsymbol{\tau}_2/18$, where $\boldsymbol{\tau}_i$ are the isospin Pauli matrices.  We have suppressed the isospin-dependent terms in \eqref{H21} by taking a trace.   Note that this contribution is $O(\rho^2)$, even though $\rho\mathcal{H}_I^{00}$ contains terms linear in $\rho$.  The linear terms do not contribute to the spin-orbit force.

Now we evaluate the contributions to $H_{LS}$ from the $\rho^2$ terms in \eqref{EH}, both of which involve sums over $N$.  We expand in terms of $\tilde{D}, A_i$ etc.\ from equation \eqref{HI} and neglect terms which are $O(M^{-2})$ or which do not involve $P$.  This leaves us with
\begin{multline*}%\label{H2a}
H_{LS}^2 =- \frac{4\mathrm{i}\rho^2\hbar}{M}P^i\sum_{N>0}\frac{\tilde{D}^{0N}\nabla_i\tilde{D}^{N0}-\nabla_i\tilde{D}^{0N}\tilde{D}^{N0}}{(E_N-E_0)^2} \\
-\frac{2\rho^2\hbar}{M\Lambda}P^i\sum_{N>0}\frac{\tilde{D}^{0N}(A_i+A_i^\dagger)^{N0}+(A_i+A^\dagger_i)^{0N}\tilde{D}^{N0}}{E_N-E_0}. 
\end{multline*}
In order to evaluate this expression one needs to know the operators $R_{ab}^{0N}$ and $R_{ab}^{N0}$ which appear in $A_i^{0N}$ etc.  Label the spins of the particles in the $E_M$ and $E_N$ eigenspaces as $(j_1,j_2)$ and $(\ell_1,\ell_2)$ respectively. Then
\begin{equation}
\label{RMN}
R_{ab}^{MN} = \kappa_a^{j_1\ell_1}\otimes\kappa_c^{j_1\ell_1}\otimes\kappa_b^{j_2\ell_2}\otimes\kappa_c^{j_2\ell_2}.
\end{equation}
Here $\kappa_a^{j\ell}$ are $(2j+1)\times(2\ell+1)$ matrices acting on spin and isospin indices, given explicitly in terms of the Clebsch-Gordan coefficients by
\begin{align}
(\kappa_1^{j\ell})_{km} &= \frac{1}{\sqrt{2}}\big(\braket{1-1\ell m|jk} - \braket{11\ell m|jk}\big)\\
(\kappa_2^{j\ell})_{km} &= \frac{\mathrm{i}}{\sqrt{2}}\big(\braket{1 1\ell m|jk} + \braket{1-1\ell m|jk}\big)\\
(\kappa_3^{j\ell})_{km} &= \braket{10\ell m|jk} 
\end{align}
where $-j\leq k\leq j$, $-\ell\leq m\leq \ell$. So, for example, $\sqrt{3}\kappa^{\frac12 \frac12}_a = -\sigma_a$. Note that $R_{ab}^{0N}$ vanishes except when $\ell_1,\ell_2=\frac12,\frac32$, corresponding to intermediate states of nucleons and delta resonances.  In evaluating $H^2_{LS}$ we once again project out the isospin-independent part by taking a trace.  The calculation takes only a few seconds on a desktop computer, and can be done by hand with effort.  The result is
 \begin{align} \label{H22}
 H_{LS}^2 = \, &\frac{\rho^2e^{-2s} }{972 \hbar^3 M X^8\Lambda^2} \boldsymbol{L}\cdot (\boldsymbol{\sigma}_1 + \boldsymbol{\sigma}_2)\Big( 64 \Lambda^4 (s^2+3s+3)^2 \nonumber \\
 &-32 \hbar^2\Lambda^2X^2(16 s^3 + 37 s^2 + 42s +21)  \nonumber
 \\ &+\hbar^4 X^4( 295 s^2  +1022 s + 727   ) \Big) .
 \end{align}
We can then simply compare our expression for $V_{SO}^{IS}$, given by the sum of \eqref{H21} and \eqref{H22}, to the isoscalar spin-orbit potential used in the phenomenological Paris potential \cite{Lacombe1980}. To plot the results, we must fix the parameters and we take those recently proposed by Lau and Manton \cite{LM2014} : $F_\pi=108\,\mathrm{MeV}$, $e=3.93$ and $m_\pi=0.7(eF_\pi/2)\approx 149\,\mathrm{MeV}$.  This fixes $C_1 = 1.815$. 

The results are shown in Figure 2 and there is good agreement between the Paris potential and $V_{SO}^{IS}$ derived from the Skyrme model at large separations. Most importantly, the sign of the spin-orbit potential is correct for long- and mid-range separations. Our approximations are valid only when $\delta\mathcal{H}$ is small compared with the energy differences $E_N-E_0$, and in particular when
\begin{equation}
\rho/X^3 < \hbar^2/\Lambda \implies X > 1.25\, \mathrm{fm}\, .
\end{equation}
The calculation is invalid below this and unreliable nearby. Hence, our poor agreement at mid-range separation may be an artefact of the approximations made. 

We now argue that a fuller treatment will improve the results. The calculation was motivated by a geometrical explanation of the spin-orbit force.  The geometrical account was based on two facts: the energetic preference for the attractive channel, and the shortness of a particular path in configuration space.  The terms in the Lagrangian \eqref{Lint} responsible for these features are the ``$D$'' and ``$A$'' terms.  The contribution $V_{SO}^{PASD}$ of these two terms to the spin-orbit potential is plotted in figure 2.  The curve is close to the Paris $V_{SO}^{IS}$ and makes the dominant negative contribution, so our geometrical explanation seems to be correct. In fact, in the dipole model axial symmetry occurs at $X=0$ while in the full model it occurs at $X\approx 1 \text{ fm}$. Hence the geometric effect should be enhanced in the full model. To understand the potential at all separations, one should carefully study the 2-skyrmion metric. The Atiyah-Manton approximation provides a promising starting point \cite{AM1992}. Note that both kinetic and potential terms are needed to obtain a negative spin-orbit potential; this explains why recent studies based only on potential terms \cite{halcrowmanton,harlandmanton} were unsuccessful.

\begin{figure}[H]
\begin{center}
\includegraphics[width=0.45\textwidth]{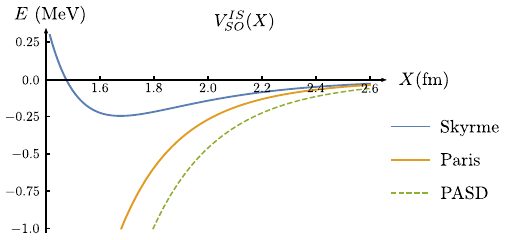}
\end{center}
\caption{A comparison between the isoscalar spin-orbit force from the Skyrme and phenomenological Paris potential.}
\label{fig:old}
\end{figure}

It is important to compare our result with earlier calculations of $V_{SO}^{IS}$ from the Skyrme model \cite{OSYKS1988,RD1988,ASW1993}.  These studies differed in two ways: they used the product approximation, rather than the dipole-dipole lagrangian, and they only worked to first order in perturbation theory.  It is now widely accepted that the product approximation is reliable at large separations (where it agrees with the dipole asymptotics) but not at small separations (since it fails to reproduce the toroidal 2-skyrmion).  Our analysis shows that at $O(\rho)$ the dipole asymptotics do not produce a spin-orbit potential, so the results of \cite{OSYKS1988,RD1988,ASW1993} must be due to short-range features of the product approximation, and are hence unreliable.  Our result is based on higher order perturbation theory, rather than first order, so is more reliable.  

In summary, we have presented a new geometrical interpretation of the spin-orbit force.  The Skyrme model, together with higher order perturbation theory, predicts a spin-orbit potential which matches the Paris potential at large separations.  We remind the reader that the phenomenological central potential is also well described by the Skyrme model using higher order perturbation theory \cite{WAH1992}.  With further development our method should allow a calculation of the complete nucleon-nucleon potential. It is also known that all skyrmions have a multipole expansion far from their center. Hence, with some modifications, these techniques can be used to model halo nuclei using the Skyrme model. While this letter has focused on the phenomenological Paris potential, our longer term ambition is to reproduce experimental scattering data directly from the Skyrme model.  The results reported here are an encouraging and important first step in this direction.

\emph{Acknowledgements} -- CJH is supported by The Leverhulme Trust as a Leverhulme Early Careers Fellow.

%\bibliographystyle{unsrtnat}
%\bibliography{nn}

%\bibliographystyle[numbers]{hunsrtnat}
%\bibliography{nn}

\bibliographystyle{utphys}
\bibliography{nn}

\end{multicols}

\end{document}